\documentclass{article}
\usepackage{geometry}
\usepackage{xcolor}
\usepackage{graphicx}
\usepackage{subcaption}
\usepackage{amsmath}
\usepackage{amssymb}
\usepackage{algorithm}
\usepackage[noend]{algpseudocode}

\definecolor{eye-caring}{RGB}{205,222,194}

\title{Distributed quantum architecture search}
\author{Haozhen Situ$^1$, Zhimin He$^2$, Shenggen Zheng$^3$, Lvzhou Li$^{4,3,}$\footnote{Corresponding author: lilvzh@mail.sysu.edu.cn}\\
{\small $^1$College of Mathematics and Informatics, South China Agricultural University, Guangzhou 510642, China}\\
{\small $^2$School of Electronic and Information Engineering, Foshan University, Foshan 528000, China}\\
{\small $^3$Quantum Science Center of Guangdong-Hong Kong-Macao Greater Bay Area (Guangdong), Shenzhen 518045, China}\\
{\small $^4$Institute of Quantum Computing and Software, School of Computer Science and Engineering,}\\
{\small Sun Yat-sen University, Guangzhou 510006, China}
}
\date{}

\begin{document}
\pagecolor{eye-caring}
\maketitle
\begin{abstract}
Variational quantum algorithms, inspired by neural networks, have become a novel approach in quantum computing. However, designing efficient parameterized quantum circuits remains a challenge. Quantum architecture search tackles this by adjusting circuit structures along with gate parameters to automatically discover high-performance circuit structures. In this study, we propose an end-to-end distributed quantum architecture search framework, where we aim to automatically design distributed quantum circuit structures for interconnected quantum processing units with specific qubit connectivity. We devise a circuit generation algorithm which incorporates TeleGate and TeleData methods to enable nonlocal gate implementation across quantum processing units. While taking into account qubit connectivity, we also incorporate qubit assignment from logical to physical qubits within our quantum architecture search framework. A two-stage progressive training-free strategy is employed to evaluate extensive circuit structures without circuit training costs. Through numerical experiments on three VQE tasks, the efficacy and efficiency of our scheme is demonstrated.
Our research into discovering efficient structures for distributed quantum circuits is crucial for near-term quantum computing where a single quantum processing unit has a limited number of qubits. Distributed quantum circuits allow for breaking down complex computations into manageable parts that can be processed across multiple quantum processing units.
\end{abstract}

\section{Introduction}\label{sec:intro}

Variational Quantum Algorithms (VQAs) \cite{cerezo2021variational} represent a novel approach to algorithm design that has emerged in recent years within the realm of quantum computing research. Drawing inspiration from machine learning methodologies, VQAs leverage Parameterized Quantum Circuits (PQCs), also known as Variational Quantum Circuits (VQCs), as analogs to neural networks. These PQCs undergo iterative updates to their gate parameters aimed at optimizing an objective function. Building upon the foundation of VQCs, a range of intriguing quantum machine learning models have been proposed. For instance, a hybrid quantum-classical neural network with deep residual learning has been proposed \cite{liang2021hybrid}. Quantum generative adversarial networks \cite{situ2020quantum,zhou2023hybrid,wang2023quantum} integrate both quantum and classical generators and discriminators, facilitating the generation of quantum or classical data. Quantum attention networks \cite{zhao2022qsan,zhao2023qksan,zhao2024gqhan} have been developed incorporating the attention mechanisms to better capture intricate interconnections among features within high-dimensional data. Quantum counterparts of denoising diffusion probabilistic models \cite{parigi2023quantum,zhang2024generative,chen2024quantum} have been devised to generate quantum state ensembles. Quantum analog of capsule networks \cite{liu2023quantum} has been developed. Deep quantum neural networks \cite{pan2023deep} and quantum convolutional neural networks \cite{herrmann2022realizing} have been realized on superconducting processors.

Drawing inspiration from Neural Architecture Search (NAS) \cite{ren2021comprehensive}, which focuses on automatically designing neural network architectures, various schemes for Quantum Architecture Search (QAS) have emerged. In QAS, not only are the gate parameters adjustable, but the structures of quantum circuits, comprising gate types and positions, are also learnable. This adaptability yields more compact PQCs tailored to specific tasks and hardware-specific qubit connectivity.

A machine learning approach has been employed to discover quantum algorithms for computing the overlap $\mathrm{Tr}(\rho\sigma)$ between two quantum states $\rho$ and $\sigma$ \cite{cincio2018learning}. Although the Swap Test serves as a standard algorithm for this purpose across various applications, the learned algorithms exhibit significantly reduced depths, with one achieving constant depth. Various aspects of QAS have been explored, including modeling the search space \cite{lu2021markovian,meng2021quantum}, refining the search strategy \cite{altares2021automatic,huang2022robust}, and enhancing the evaluation methods \cite{zhang2021neural,du2022quantum}.

In this work, our focus lies within the domain of distributed quantum computing \cite{caleffi2022distributed}. By interconnecting and coordinating multiple small-scale Quantum Processing Units (QPUs), we aim to leverage a greater number of qubits to tackle larger-scale problems, thereby presenting a promising architecture for near-term quantum computing. Our objective is to automatically devise distributed quantum circuit structures tailored for specific VQA tasks. To the best of our knowledge, this represents the first endeavor employing QAS to design distributed quantum circuits.

Concretely, we first propose a novel circuit generation algorithm. This algorithm begins by transforming the graph describing the distributed system into a virtual connectivity graph. Subsequently, it iteratively samples gate types and positions randomly. Two distinct methods facilitating the implementation of nonlocal gates across various QPUs are integrated into our algorithm. Besides taking into account the qubit connectivity of QPUs, our distributed QAS framework also incorporates the qubit assignment process into its workflow, which is often overlooked in previous QAS literature. To circumvent the training costs associated with VQCs and enable exploration across a broader spectrum of circuit structures, we adopt the training-free evaluation methodology \cite{he2024training-free}. This methodology streamlines vast circuit structures through two stages, utilizing a path-based proxy and an expressibility-based proxy as filters. We evaluate our distributed QAS framework through three VQE tasks, demonstrating both the efficacy and efficiency of our approach through numerical experimentation.

The proposed scheme, which adjusts both circuit structures and gate parameters, shares similarities with the VAns (Variable Ansatz) method \cite{bilkis2023semi} and the ADAPT-VQE (Adaptive Derivative-Assembled Pseudo-Trotter ansatz Variational Quantum Eigensolver) algorithm \cite{grimsley2019adaptive}. The VAns method applies a set of rules to both add and remove quantum gates from the circuit in an informed manner during the optimization, while the ADAPT-VQE algorithm incrementally expands the circuit by adding gates that implement fermionic operators selected from a pool of single and double excitation operators. Compared to these methods, the proposed distributed QAS framework offers a distinct approach. Instead of using a nested optimization loop where the outer loop optimizes the circuit structure and the inner loop optimizes the gate parameters, the proposed distributed QAS employs training-free proxies to evaluate circuit structures without optimizing the gate parameters. Only a filtered selection of promising candidate circuit structures then undergo the gate parameter optimization process. Additionally, the proposed scheme features the integration of qubit assignment and the implementation of nonlocal gates, resulting in a comprehensive end-to-end distributed QAS framework.

The remainder of this article is organized as follows. In Section \ref{sec:related}, some related work on QAS are reviewed. Then, in Section \ref{sec:DQAS}, we present our distributed QAS framework including qubit assignment, methods for nonlocal gate implementation, virtual connective graph construction, distributed circuit generation, and the search strategy. Section \ref{sec:eval} is dedicated to the evaluation of our proposed framework across three VQE tasks. Section \ref{sec:discussion} discusses some issues related to our framework. Finally, Section \ref{sec:conclusion} concludes this article.

\section{Related work}\label{sec:related}

A neural network based predictor has been used as the evaluation policy for QAS \cite{zhang2021neural}. Rather than training quantum circuits to assess their performance, this method trains a neural predictor to directly gauge the performance of quantum circuits using only their structures. This predictor is then integrated into the QAS workflow to accelerate the search process. A graph self-supervised methodology was introduced to improve predictor based QAS \cite{he2023gsqas}. A graph encoder is pre-trained using a well-designed pretext task on a large number of unlabeled quantum circuits, aiming to generate meaningful representations of quantum circuits. Subsequently, the downstream predictor is trained on a small set of quantum circuit representations paired with their labels. Once the encoder is trained, it becomes applicable to various downstream tasks.

A quantum neuroevolution algorithm was introduced to autonomously find near-optimal quantum neural networks for different machine learning tasks \cite{lu2021markovian}. This algorithm establishes a one-to-one mapping between quantum circuits and directed graphs, reducing the problem of finding the appropriate gate sequences to a task of searching suitable paths in the corresponding graph as a Markovian process. NSGA-II (Non-Sorted Genetic Algorithm II) was employed to automatically generate optimal ad-hoc ansatz for classification tasks utilizing quantum support vector machines \cite{altares2021automatic}. This multiobjective genetic algorithm enables the simultaneous maximization of accuracy and minimization of ansatz size. A genome-length-adjustable evolutionary algorithm was utilized to design a robust VQA circuit that is optimized over variations of both circuit ansatz and gate parameters, without any prior assumptions on circuit structure or depth \cite{huang2022robust}.
EQNAS, an evolutionary QAS algorithm designed for image classification, was introduced \cite{li2023eqnas}. This algorithm initiates a quantum population following quantum image encoding, and further refines it through the application of quantum rotation gates and entirety interference crossover operations.

A quantum circuit architecture optimization algorithm leveraging Monte Carlo Tree (MCT) search was proposed \cite{meng2021quantum}. This algorithm first models the search space with an MCT that can be regarded as a supernet. During MCT training, the weight sharing strategy is utilized to reduce computation cost. Training results are stored in MCT nodes for future decisions, and hierarchical node selection is applied to obtain an optimal ansatz. An algorithmic framework, combing nested MCT search with the combinatorial multi-armed bandit model, was introduced for the automatic design of quantum circuits \cite{wang2023automated}.

A QAS scheme incorporating supernet and weight sharing was introduced \cite{du2022quantum}. This method establishes a supernet defining the ansatz pool, parameterizing each ansatz via weight sharing strategy. Subsequently, it iteratively samples an ansatz from the pool and optimize its parameters. Following evaluation across a number of ansatz, the top-performing candidate is selected and finetuned with few iterations.
QuantumNAS is a comprehensive framework for noise-adaptive co-search of variational circuits and qubit mapping \cite{wang2022quantumnas}. Initially, a supercircuit is constructed and trained through iteratively sampling and updating the subcircuits. Then, an evolutionary co-search of subcircuit and its qubit mapping is deployed. Finally, iterative gate pruning and finetuning procedures are executed to eliminate redundant gates.

A general framework of differentiable QAS was proposed \cite{zhang2022differentiable}. This approach involves the relaxation of the discrete search space of quantum circuit structures onto a continuous and differentiable domain, enabling optimization through gradient descent. QuantumDARTS is a differentiable QAS based on Gumbel-Softmax \cite{wu2023quantumdarts}. This algorithm distinguishes itself from existing methods that typically demand extensive circuit sampling and evaluation. It introduces a micro search strategy to infer the subcircuit structure from a small-scale problem and then transfer it to a large-scale context. A gradient-based QAS algorithm enhanced with meta-learning was introduced \cite{he2022quantum}. This approach learns good initialization heuristics of the architecture, along with the meta-parameters of quantum gates from a number of training tasks. By doing so, it enables the algorithm to swiftly adjust to new tasks with minimal gradient updates, facilitating fast learning on new tasks.

QAS algorithms require calculating the performances for a large number of circuits during the search process, which incurs substantial computational costs due to the iterative updating of gate parameters. The predictor-based approach alleviates the training cost of PQCs by using a predictor to approximate the circuit performance. However, the predictor is trained using a supervised learning approach, which necessitates calculating the ground-truth performances for the circuits in the training set. The quantity of circuit-performance pairs is crucial for the predictor's generalization ability. In the weight sharing approach, the performance of a sub-circuit is estimated using parameters inherited from the super-circuit. Nevertheless, there is no guarantee that the performance of a quantum circuit with inherited parameters will strongly correlate with the performance when trained individually.

A training-free QAS approach, utilizing two proxies to rank quantum circuits in place of the expensive circuit training, was introduced \cite{he2024training-free}. Initially, directed acyclic graphs are utilized for circuit representation, and a zero-cost proxy based on the number of paths in the directed acyclic graph effectively filters out a substantial portion of unpromising circuits. Subsequently, an expressibility-based proxy, finely reflecting circuit performance, is employed to identify high-performance circuits from the remaining candidates.
However, like most QAS literature, qubit assignment is not considered.

To fill the gap in automatically designing quantum circuits for distributed quantum computing, we propose a distributed QAS framework, which incorporates the latest training-free QAS approach to avoid expensive circuit training cost. The qubit assignment overlooked by previous literature is integrated into the optimization process, to better utilize the qubit topology property of the quantum device. Additionally, we incorporate two methods for implementing nonlocal gates, TeleGate and TeleData, resulting in a flexible distributed QAS framework.

\section{Distributed QAS}\label{sec:DQAS}

\begin{figure}
	\centering
	\includegraphics[width = 11cm]{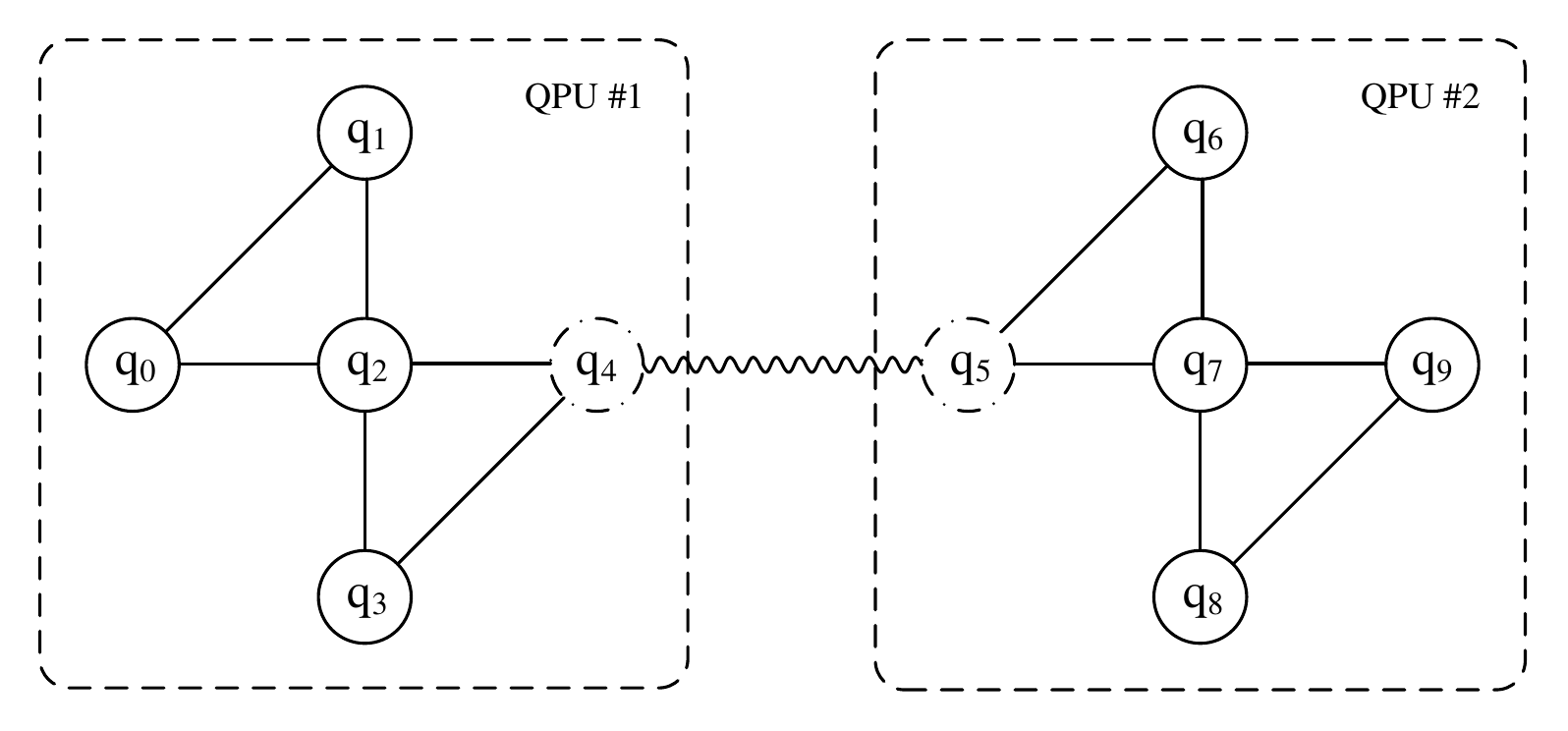}	
	\caption{Two IBM Yorktown quantum processors are interconnected via a quantum link. Circles with solid lines represent data qubits, while circles with dash-dotted lines represent communication qubits. Solid lines denote local couplings, while the wavy line represents the quantum link for distributing pairs of entangled qubits to two communication qubits.}
	\label{fig:device}
\end{figure}

A distributed quantum computing system comprises multiple QPUs interconnected by quantum links. It can be characterized by a graph $G=(Q, E, L)$, where $Q$ represents the set of qubits, $E$ represents the set of coupling edges between qubits on the same QPU, and $L$ represents the set of quantum links, each connecting two communication qubits from different QPUs. There are two types of qubits on QPUs, the data qubits $Q_d$ and communication qubits $Q_c$. The data qubits are dedicated to computation, while the communication qubits are reserved for facilitating nonlocal operations, which will be elaborated later. Fig.\ \ref{fig:device} depicts a distributed quantum computing system where two 5-qubit QPUs are interconnected by a quantum link. The data qubit set $Q_d$ consists of $\{q_0, q_1, q_2, q_3, q_6, q_7, q_8, q_9\}$, while the communication qubit set $Q_c$ consists of $\{q_4, q_5\}$. The total number of data qubits available for assignment of logical qubits is 8, implying a maximum problem size of  8 qubits.

Without loss of generality, we assume the edges between qubits are undirected, enabling a $CNOT$ gate to be applied on a pair of adjacent qubits regardless of which one serves as the control qubit. Additionally, we assumes that each data qubit is connected to at most one communication qubit.

\subsection{Qubit assignment}
As the connectivity of physical qubits depends on specific QPU, certain qubits exhibit more connections to others, while some have less. Thus, a well-suited qubit assignment becomes crucial for effectively addressing computation tasks. Qubit assignment is a mapping from logical qubits to data qubits, which can be defined as
\begin{align}
f_{q}: Q_{l} \longrightarrow Q_{d},
\end{align}
where $Q_l$ and $Q_d$ denote logical qubits and data qubits, respectively. If there are more data qubits available than logical qubits required, some data qubits are reserved to accommodate teleported qubits. These reserved qubits are termed as empty qubits.

It's worth noting that previous research on QAS has rarely included qubit assignment $f_q$ as a part of the optimization process, despite considering qubit connectivity. In this study, we integrate the optimization of $f_q$ into our QAS framework to learn an optimal $f_q$.

\subsection{Nonlocal gate implementation}
Without the use of quantum links, only single-qubit and local two-qubit gates can be executed, while a nonlocal two-qubit gate involving two qubits from different QPUs is not feasible. In this work, we assume the native gate set includes single-qubit $U$ gates, two-qubit $CNOT$ gates, and $SWAP$ gates, defined as follows:
\begin{align}
U(\theta, \phi, \lambda) = \left[
\begin{array}{cc}
  \cos(\theta /2) & -e^{i\lambda}\sin(\theta/2) \\
  e^{i\phi}\sin(\theta/2) & e^{i(\phi+\lambda)}\cos(\theta/2) \\
  \end{array}
\right],\ \ \
CNOT = \left[
\begin{array}{cccc}
  1 & 0 & 0 & 0 \\
  0 & 1 & 0 & 0 \\
  0 & 0 & 0 & 1 \\
  0 & 0 & 1 & 0 \\
  \end{array}
\right],\ \ \
SWAP = \left[
\begin{array}{cccc}
  1 & 0 & 0 & 0 \\
  0 & 0 & 1 & 0 \\
  0 & 1 & 0 & 0 \\
  0 & 0 & 0 & 1 \\
  \end{array}
\right].
\end{align}
We use superscripts and subscripts to denote the control and target qubits of $CNOT$ gates. For instance, $CNOT^a_b$ indicates that qubit $a$ serves as the control and qubit $b$ serves as the target.

In quantum distributed computing, two widely used methods for implementing nonlocal gates are the TeleGate method and the TeleData method \cite{caleffi2022distributed}. Both methods leverage shared entangled qubits and classical communication to realize nonlocal gates.

\subsubsection{TeleGate method}\label{sec:TeleGate}

We first describe the TeleGate method. Suppose the state of data qubits is represented by $|\psi\rangle$. By treating all the qubits as a single composite system, we can express its state as
\begin{align}
|\psi\rangle(|00\rangle+|11\rangle)_{ab} = (|0\rangle_c|\psi_0\rangle + |1\rangle_c|\psi_1\rangle)(|00\rangle+|11\rangle)_{ab},
\end{align}
where the subscript $c$ denotes the control qubit of the nonlocal $CNOT$ gate, while subscripts $a$ and $b$ denote the communication qubits sharing a Bell state. Note that $|\psi_0\rangle$ and $|\psi_1\rangle$ are unnormalized for brevity. Qubits $c$ and $a$ are situated within one QPU and connected by an edge, while qubit $b$ is located on another QPU.

To implement nonlocal $CNOT$ gates, the TeleGate method initially employs a ``cat-entangler'' primitive operation \cite{yimsiriwattana2004generalized}, which transforms the quantum state of the entire system to
\begin{align}\label{eq:cat}
|00\rangle_{cb}|\psi_0\rangle + |11\rangle_{cb}|\psi_1\rangle.
\end{align}
The details of how this transformation works are given in Appendix A.
We can see that after the cat-entangler, the control qubit $c$ on one QPU and the communication qubit $b$ on another QPU become entangled. Consequently, a local $CNOT$ gate controlled by qubit $b$ is equivalent to a nonlocal $CNOT$ gate controlled by qubit $c$. For ease of explanation later on, we refer to qubit $c$ as being in ``control mode''.

In the example depicted in Fig.\ \ref{fig:device}, if a cat-entangler operation is performed on $q_2q_4q_5$, the resulting state becomes
\begin{align}
|00\rangle_{q_2q_5}|\psi_0\rangle + |11\rangle_{q_2q_5}|\psi_1\rangle.
\end{align}
When using $q_5$ as the control qubit, local $CNOT^{q_5}_{q_6}$ is equivalent to nonlocal $CNOT^{q_2}_{q_6}$. Analogously, local $CNOT^{q_5}_{q_7}$ is equivalent to nonlocal $CNOT^{q_2}_{q_7}$. In this way, two nonlocal $CNOT$ gates become possible. At the same time, qubit $q_2$ can still participate in local $CNOT$ gates, such as $CNOT^{q_2}_{q_0}$, $CNOT^{q_2}_{q_1}$, and $CNOT^{q_2}_{q_3}$.

The nonlocal gates mentioned above remains feasible as long as qubit $c$ remains in control mode, but it cannot act as the target of other gates. To transition qubit $c$ out of control mode, the ``cat-disentangler'' primitive operation \cite{yimsiriwattana2004generalized} is required. The cat-disentangler transforms Eq. \ref{eq:cat} to
\begin{align}
|0\rangle_{c}|\psi_0\rangle + |1\rangle_{c}|\psi_1\rangle.
\end{align}
The details of how this transformation works are given in Appendix A.
After the cat-disentangler operation, qubit $c$ transitions out of control mode and can participate in all local gates. Since communication qubits $a$ and $b$ have been measured, the shared entanglement is depleted. A new pair of entangled qubits must be distributed among qubits $a$ and $b$. We say that the cat-entangler and the cat-disentangler cost 1 ebit.

\begin{figure*}
	\centering
    \includegraphics[width = 8cm]{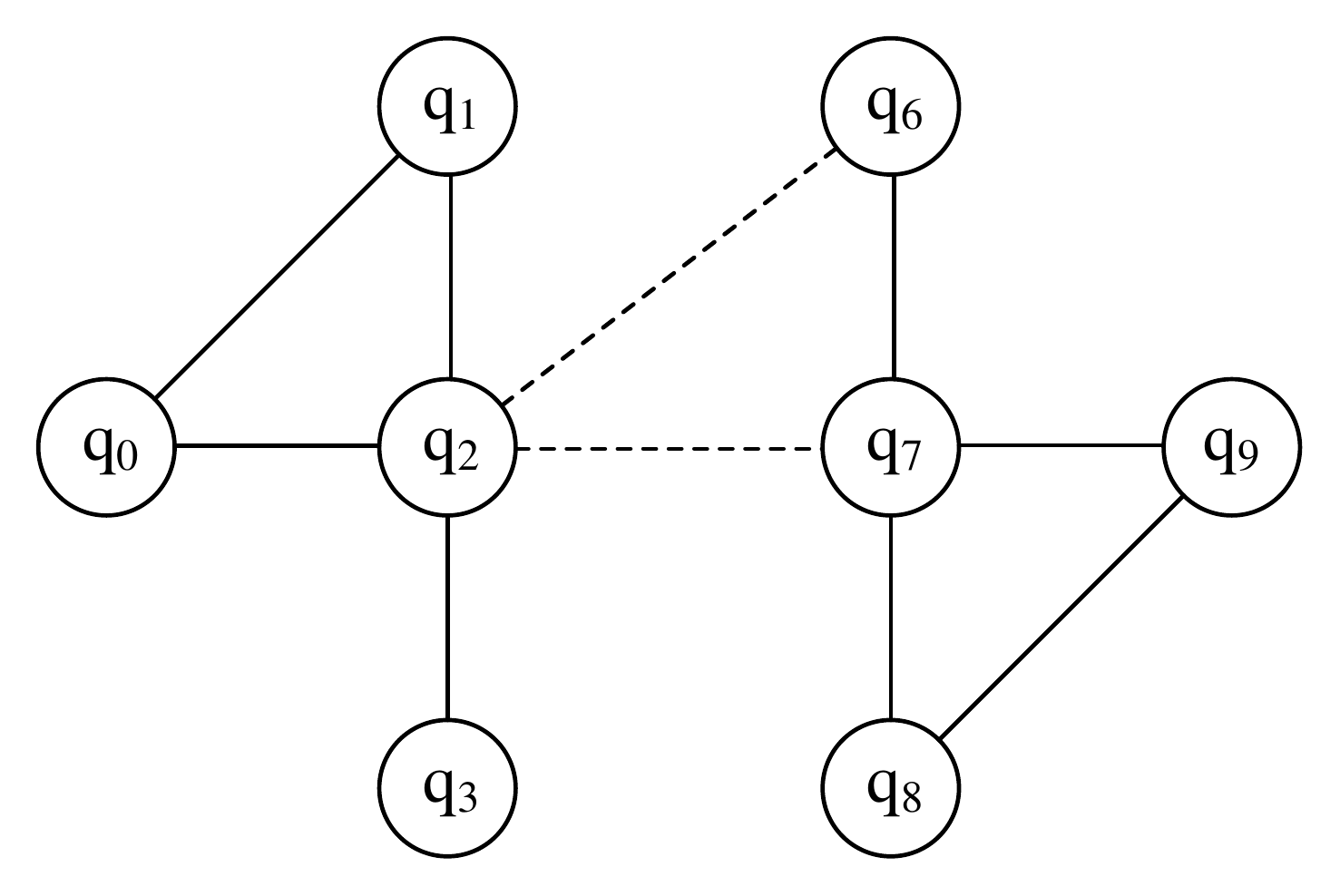}
    \caption{The virtual connectivity graph is derived from the distributed architecture depicted in Fig.\ \ref{fig:device}. Local two-qubit gates can be directed performed on qubits connected by solid lines. The two dash lines represent virtual edges added through the cat-entangler primitive operation on qubits $q_2q_4q_5$.}
	\label{fig:VCG}
\end{figure*}

In order to hide the intricacies of cat-entangler and cat-disentangler operations within our QAS framework, we can ignore the presence of communication qubits. We dynamically add or remove ``virtual edges'' whenever these primitive operations occur. Fig.\ \ref{fig:VCG} depicts the virtual connectivity graph corresponding to the data qubits in Fig.\ \ref{fig:device}. When a cat-entangler operation is performed on $q_2q_4q_5$, virtual edges $(q_2,q_6)$ and $(q_2,q_7)$ are inserted with $q_2$ entering control mode. With the presence of these virtual edges and the knowledge that $q_2$ is in control mode, two nonlocal $CNOT$ gates $CNOT^2_6$ and $CNOT^2_7$ become feasible. Similarly, a cat-entangler operatation on $q_3q_4q_5$ results in the insertion of virtual edges $(q_3,q_6)$ and $(q_3,q_7)$, with $q_3$ entering control mode. A cat-entangler operation on $q_6q_5q_4$ leads to the insertion of virtual edges $(q_6,q_2)$ and $(q_6,q_3)$, with $q_6$ entering control mode. A cat-entangler operated on $q_7q_5q_4$ results in the insertion of virtual edges $(q_7,q_2)$ and $(q_7,q_3)$, with $q_7$ entering control mode.  When a cat-disentangler operation occurs, the associated virtual edges are removed, and the corresponding qubit in control mode transitions out of control mode.

\subsubsection{TeleData method}
When there are more data qubits available than logic qubits required for the computation task, empty qubits are present. To transmit a qubit from one QPU to the empty qubit of another QPU, the teleportation protocol \cite{bennett1993teleporting} is employed, as direct qubit transmission is susceptible to decoherence. Teleportation is a well-known quantum protocol, and we'll omit its detailed workings here. In essence, a data qubit adjacent to a communication qubit of QPU 1 and an empty qubit adjacent to a communication qubit of QPU 2 can be swapped if the two communication qubits share a pair of entangled qubits.
In the example depicted in Fig.\ \ref{fig:device}, in order to perform a nonlocal $CNOT$ gate on $q_2q_8$, we can teleport $q_2$ to $q_5$, swap $q_5$ and $q_7$ (which must be empty), and then perform a $CNOT$ gate on $q_7q_8$.

The TeleData method incurs the same cost of 1 ebit as the TeleGate method. If the communication qubits required for teleportation are already occupied by the cat-entangler, no entangled qubits are available for teleportation. In such cases, the cat-disentangler can be invoked to release the communication qubits, allowing for the distribution of a new pair of entangled qubits between them.

\subsection{Virtual connectivity graph}

Before generating random circuits, we first deduce a virtual connectivity graph from the graph description of the distributed quantum computing system, denoted as $G=(Q,E,L)$. The virtual connectivity graph only includes data qubits $Q_d\subset Q$, excluding edges involving communication qubits. The permissable positions of local gates are defined by
\begin{align}\label{eq:VCG1}
S_{l}\leftarrow E - \{(a,b)\in E| a \notin Q_d \vee b \notin Q_d\}.
\end{align}
$SWAP$ gates that do not alter qubit connectivity are prohibited. For instance, $SWAP(q_0,q_1)$ in Fig.\ \ref{fig:VCG} is not allowed because neighbours of $q_0$ and $q_1$ remain unchanged after the $SWAP$ gate. The permissable positions of $SWAP$ gates are defined by
\begin{align}\label{eq:VCG2}
S_s\leftarrow \{(a,b)\in S_l| R(a) -\{b\} \neq R(b)-\{a\}\},
\end{align}
where $R(x)=\{y|(x,y)\in S_l \}$. The TeleGate method results in adding virtual edges to the virtual connectivity graph. The permissable virtual edges are defined by
\begin{align}\label{eq:VCG3}
S_{tg}\leftarrow \cup_{(a,b)\in L} R'(a) \times R'(b),
\end{align}
where $R'(x) = \{y | (x,y)\in E\}$. The permissable positions of nonlocal gates implemented through the TeleData method are defined by
\begin{align}\label{eq:VCG4}
S_{td}\leftarrow \cup_{(a,b)\in L} ((R'(R'(a)) -\{a\})  \times R'(b)) \cup (R'(a) \times (R'(R'(b)) -\{b\})).
\end{align}

\subsection{Circuit generation}
In QAS, a large number of circuit structures are explored. The algorithm for generating random distributed quantum circuit structures is provided in Algorithm \ref{alg:sampling}. Given the graph description of the distributed quantum computing system, our algorithm initially identifies the permissible positions of local gates and nonlocal gates. It then proceeds iteratively, adding gates by randomly selecting their types and positions. Our algorithm ensures that empty qubits cannot be the objectives of $U$ and $CNOT$ gates, and at least one of the participating qubits in a $SWAP$ gate is nonempty. The proportion of local and nonlocal two-qubit gates is predefined. The TeleGate and TeleData procedures are invoked when a nonlocal gate needs to be added.

Three categories of distributed circuits can be generated: (1) circuits with nonlocal gates implemented through the TeleGate method, (2) circuits with nonlocal gates implemented through the TeleData method, and (3) circuits with nonlocal gates implemented through either TeleGate or TeleData.

\begin{algorithm}[htp]
\renewcommand{\algorithmicrequire}{\textbf{Input:}}
\renewcommand{\algorithmicensure}{\textbf{Output:}}
\setcounter{algorithm}{0}
\caption{Generation of distributed quantum circuits}
\label{alg:sampling}
\begin{algorithmic}[1]
\Require
$G=(Q,E,L)$: a graph description of the distributed quantum computing system;
$N_g$: the number of gates in a circuit;
$P_{g}$: the distribution of gate types;
$p_{nl}$: the probability of nonlocal gates;
$M$: the method to implement nonlocal gates.
\Ensure $C$: a distributed quantum circuit;
$N_e$: the number of ebits
\State Deduce the virtual connectivity graph according to Eqs. \ref{eq:VCG1},\ref{eq:VCG2},\ref{eq:VCG3},\ref{eq:VCG4}.
\State $C\leftarrow\emptyset$
\State $N_e \leftarrow 0$
\While{$|C| < N_g$}
    \State select a random gate type $g\in\{U, CNOT, SWAP\}$ according to $P_{g}$
    \If {g = $U$}
        \State select a random qubit $q$ from non-empty qubits
        \If {$\mathrm{redundant}(g, q)$ = False}
            \If {$\mathrm{control\_mode}(q)$ = True}
                \State remove virtual edges related to $q$
                \State $\mathrm{control\_mode}(q) \leftarrow$ False
            \EndIf
            \State $C\leftarrow C \cup \{(g,q)\}$
        \EndIf
    \ElsIf {g = $CNOT$}
        \State pick a random value $a\in(0,1)$
        \If {$a>p_{nl}$}
            \State randomly choose the control $c$ and the target $t$ from $S_l$ satisfying neither $c$ nor $t$ is empty
            \If {$\mathrm{redundant}(g, c, t)$ = False}
                \If {$\mathrm{control\_mode}(t)$ = True}
                    \State remove virtual edges related to $t$
                    \State $\mathrm{control\_mode}(t) \leftarrow$ False
                \EndIf
                \State $C\leftarrow C \cup \{(g,c,t)\}$
            \EndIf
        \Else
            \State call NonlocalGate (procedure \ref{proc:NonlocalGate})
        \EndIf
    \ElsIf {g = $SWAP$}
        \State select a random position $(a,b)$ from $S_s$ satisfying either $a$ or $b$ is non-empty
        \If {$\mathrm{redundant}(g, a, b)$ = False}
            \If {$\mathrm{control\_mode}(a) = \mathrm{True}$}
                \State remove virtual edges related to $a$
                \State $\mathrm{control\_mode}(a) \leftarrow$ False
            \EndIf
            \If {$\mathrm{control\_mode}(b) = \mathrm{True}$}
                \State remove virtual edges related to $b$
                \State $\mathrm{control\_mode}(b) \leftarrow$ False
            \EndIf
            \State $C\leftarrow C \cup \{(g,a,b)\}$
        \EndIf
    \EndIf
\EndWhile
\Return $C$, $N_e$
\end{algorithmic}
\end{algorithm}

\begin{algorithm}[t]
\floatname{algorithm}{Procedure}
\caption{NonlocalGate}
\label{proc:NonlocalGate}
\begin{algorithmic}[1]
\If {$M=$TeleGate}
    \State $S_{tg}' \leftarrow \{(a,b)\in S_{tg} | \mathrm{nonempty}(a) = \mathrm{True} \wedge  \mathrm{nonempty}(b) = \mathrm{True}\}$
    \If {$S_{tg}' \neq \emptyset$}
        \State randomly select the control $c$ and the target $t$ from $S_{tg}'$
        \If {$\mathrm{redundant}(g, c, t)$ = False}
            \If {$\mathrm{control\_mode}(c)$ = True}
                \State $C\leftarrow C \cup \{(g,c,t)\}$
            \Else
                \State call TeleGate (procedure \ref{proc:TeleGate})
            \EndIf
        \EndIf
    \EndIf
\ElsIf {$M=$TeleData}
    \State $S_{td}' \leftarrow \{(a,b)\in S_{td} | \mathrm{nonempty}(a) = \mathrm{True} \wedge  \mathrm{nonempty}(b) = \mathrm{True}\}$
    \If {$S_{td}' \neq \emptyset$}
        \State randomly select the control $c$ and the target $t$ from $S_{td}'$
        \If {$\mathrm{redundant}(g, c, t)$ = False}
            \If {empty qubit to facilitate teleportation exists}
                \State call TeleData (procedure \ref{proc:TeleData})
            \EndIf
        \EndIf
    \EndIf
\ElsIf {$M=$Both}
    \State $S' \leftarrow \{(a,b)\in S_{tg} \cup S_{td} | \mathrm{nonempty}(a) = \mathrm{True} \wedge  \mathrm{nonempty}(b) = \mathrm{True}\}$
    \If {$S' \neq \emptyset$}
        \State randomly select the control $c$ and the target $t$ from $S'$
        \If {$\mathrm{redundant}(g, c, t)$ = False}
            \If {$g\in S_{tg}$}
                \If {$\mathrm{control\_mode}(c)$ = True}
                    \State $C\leftarrow C \cup \{(g,c,t)\}$
                \Else
                    \If {$g \in S_{td}\ \wedge $ empty qubit to facilitate teleportation exists}
                        \State call TeleGate (procedure \ref{proc:TeleGate}) or TeleData (procedure \ref{proc:TeleData}) with equal probability
                    \Else
                        \State call TeleGate (procedure \ref{proc:TeleGate})
                    \EndIf
                \EndIf
            \Else
                \If {empty qubit to facilitate teleportation exists}
                    \State call TeleData (procedure \ref{proc:TeleData})
                \EndIf
            \EndIf
        \EndIf
    \EndIf
\EndIf
\end{algorithmic}
\end{algorithm}

\begin{algorithm}[t]
\floatname{algorithm}{Procedure}
\caption{TeleGate}
\label{proc:TeleGate}
\begin{algorithmic}[1]
\If {required quantum link in use}
    \State remove associated virtual edges to release the link
    \State let the corresponding control qubit exit the control mode
\EndIf
\State add virtual edges related to $c$
\State $\mathrm{control\_mode}(c) \leftarrow$ True
\State $C\leftarrow C \cup \{(g,c,t)\}$
\State $N_e\leftarrow N_e+1$
\end{algorithmic}
\end{algorithm}

\begin{algorithm}[t]
\floatname{algorithm}{Procedure}
\caption{TeleData}
\label{proc:TeleData}
\begin{algorithmic}[1]
\State from qubits $c$ and $t$, select the qubit to be teleported and denote it as $a$
\If {required quantum link in use}
    \State remove associated virtual edges to release the link
    \State let the corresponding control qubit exit the control mode
\EndIf
\If {$\mathrm{control\_mode}(t)$ = True}
    \State remove virtual edges related to $t$
    \State $\mathrm{control\_mode}(t) \leftarrow$ False
\EndIf
\State teleport qubit $a$
\State $C\leftarrow C \cup \{(g,c,t)\}$
\State $N_e\leftarrow N_e+1$
\end{algorithmic}
\end{algorithm}

\begin{figure}
	\centering
	\includegraphics[width = 13cm]{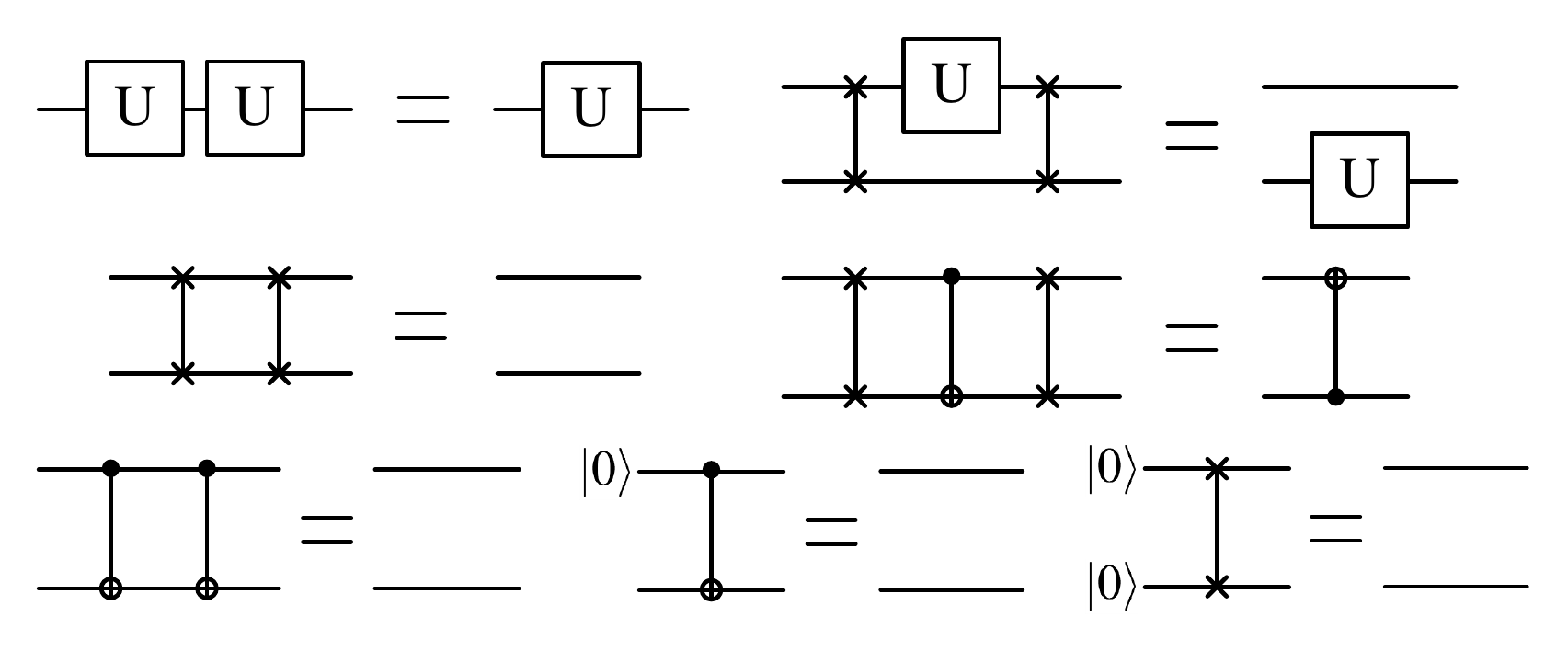}	
	\caption{Examples of redundant gates.}
	\label{fig:redundancy}
\end{figure}

Before adding a gate, some requirements are checked to prevent redundancy. Some examples of redundant gates are illustrated in Fig.\ \ref{fig:redundancy}. In order to generate more efficient circuits, we ensure that: (1) two consecutive gates cannot be of the same type because two $U$ gates can be combined into one, and two $CNOT$ gates with the same control and target cancel each other out. (2) After $SWAP(a,b)$, another $SWAP(a,b)$ is permitted if $a$ or $b$ has participated in a two-qubit gate (but not the same two-qubit gate), or has been involved in the TeleData procedure. (3) If the initial state of the circuit is $|0\rangle$, a $CNOT$ gate cannot be the first gate of its control qubit. Additionally, $SWAP(a,b)$ is not allowed if $a$ and $b$ have not been affected by other gates except when one of them is an empty qubit.

\subsection{Search strategy}
In this work, we employ a two-stage progressive strategy free from training requirements \cite{he2024training-free}, to navigate the expansive search space effectively. This strategy utilizes two training-free proxies to evaluate the performance of numerous circuits efficiently. It filters out unpromising circuits, retaining only a select few candidates. Subsequently, these candidate circuits are trained on specific VQA task to ascertain the optimal circuit.

Following the generation of $K_a$ distributed quantum circuits, the number of paths between the input and output nodes of the directed acyclic graph representation of each circuit is calculated. The circuits are then sorted in descending order based on their number of paths, and the top $K_p$ circuits are selected. Because of its low computational cost and its capability to capture the topological complexity of quantum circuits, the path-based proxy can effectively serve as a preliminary filtering mechanism to eliminate poor-performance circuits.

Next, the expressibility \cite{sim2019expressibility} of each of the $K_p$ circuits is evaluated as follows
\begin{align}
\mathcal{E}(C) = D_{\mathrm{KL}}(P_C(F)||P_{\mathrm{Haar}}(F)),
\end{align}
where $P_C(F)$ represents the distribution of $F=|\langle 0|V^\dagger(\theta)|V(\theta')|0\rangle|^2$ generated by sampling random gate parameters $\theta$ and $\theta'$, $P_{\mathrm{Haar}}(F)$ represents the distribution of $F=|\langle\psi|\psi'\rangle|^2$ with $\psi$ and $\psi'$ being Haar random states, and $D_{\mathrm{KL}}$ denotes the Kullback-Leibler divergence. A smaller expressibility value $\mathcal{E}(C)$ indicates better expressibility of the circuit $C$. These circuits are once again sorted in ascending order of their expressibility values, and the top $K_e$ circuits are chosen to constitute the candidate circuit set. Expressibility reflects the quantum circuits' capacity to uniformly reach the entire Hilbert space, making it a more precise filtering proxy. Finally, a small subset of candidate circuits that survive the two-stage filtering process undergo tailored training specific to the problem at hand.

\section{Evaluation}\label{sec:eval}

We evaluate our distributed quantum architecture search framework using three 6-qubit VQE tasks: the BeH$_2$ molecule, the Heisenberg model and the Transverse-Field Ising Model (TFIM). The Hamiltonian for BeH$_2$ at its lowest-energy interatomic distance (bond distance) is constructed according to Ref. \cite{kandala2017hardware}. The Hamiltonian of the Heisenberg model with periodic boundary condition is given by $H_{\mathrm{Heisenberg}} = \sum_{i} X_i X_{i+1} + Y_i Y_{i+1}+ Z_i Z_{i+1} + Z_i$. The Hamiltonian of the TFIM with periodic boundary condition is $H_{\mathrm{TFIM}} = \sum_{i} Z_i Z_{i+1} + X_i$. The goal of VQE tasks is to determine the ground state energy of these Hamiltonians, denoted as $E_{g} = \min_{\theta}\langle 0|V^{\dagger}(\theta) H V(\theta)|0\rangle$, where $V(\theta)$ represents a variational quantum circuit with gate parameters denoted by $\theta$. A solution is considered optimal if it falls within the chemical accuracy threshold of 0.0016 of the ground state energy.

The generated circuits adhere strictly to the topological constraints illustrated in Fig.\ \ref{fig:device}. To enhance the diversity of generated circuits, the distribution of gate types $P_g$ for each circuit is chosen randomly from three predefined distributions: $(0.4, 0.2, 0.4)$, $(0.5, 0.25, 0.25)$, and $(0.6, 0.3, 0.1)$, where the three probabilities in each distribution correspond to $U$, $CNOT$ and $SWAP$ gates, respectively. Additionally, the probability of nonlocal gates $p_{nl}$ for each circuit is randomly selected from $\{0.1,0.2,0.3,0.4\}$. We maintain that the number of $U$ gates is not less than the number of $CNOT$ gates, and the number of nonlocal gates does not exceed the number of local gates.

In the numerical simulations, we set the values of $K_a$, $K_p$, and $K_e$ to $100000$, $10000$, and $1000$, respectively. The hyperparameters $K_a$, $K_p$, and $K_e$ determine the number of circuits whose number of paths, expressibility, and performance will be calculated, respectively. We choose their values by balancing the need for a sufficient number of samples with acceptable computation time. The $K_e$ circuits, sorted based on their expressibility, serve as candidate circuits for the VQE tasks. These candidate circuits are trained individually, following the order of expressibility from better to worse. We refer to the training of each candidate circuit as a query. During each query, the candidate circuit undergoes training with 10 random initializations of parameters until convergence, achievement of chemical accuracy, or reaching the maximum number of iterations 10000. The minimal energy found within 10 runs is regarded as the performance of a PQC. For simulation and training of quantum circuits, we utilize the TensorCircuit framework \cite{zhang2023tensorcircuit}. We employ the Adam optimizer with a learning rate of 0.01 to optimize the parameters of the circuits.

\begin{figure}[ht]
    \centering
    \begin{subfigure}[b]{0.48\textwidth}
        \centering
        \includegraphics[width=\textwidth]{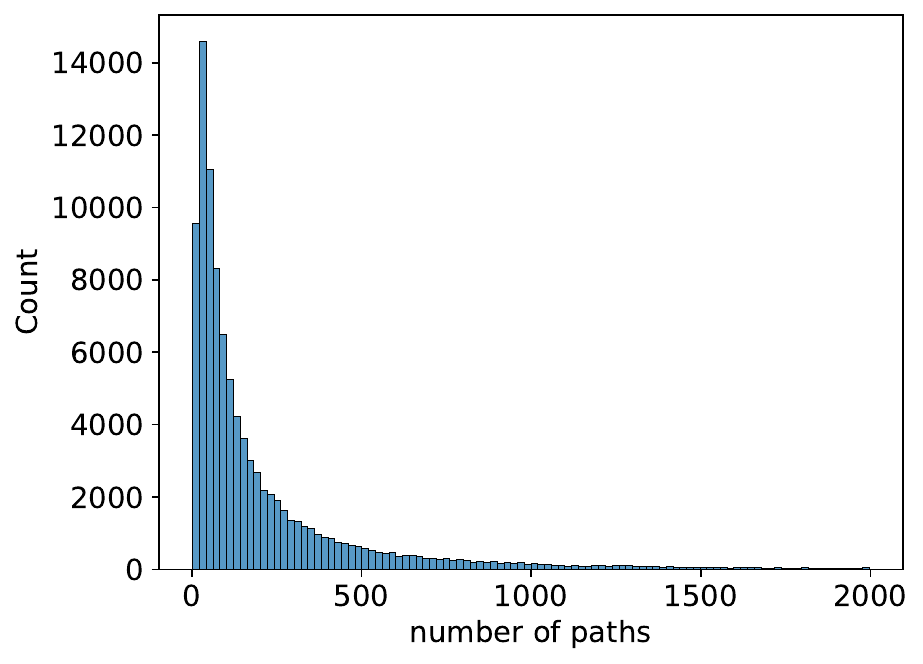}
        \caption{}
    \end{subfigure}
    \hfill
    \begin{subfigure}[b]{0.48\textwidth}
        \centering
        \includegraphics[width=\textwidth]{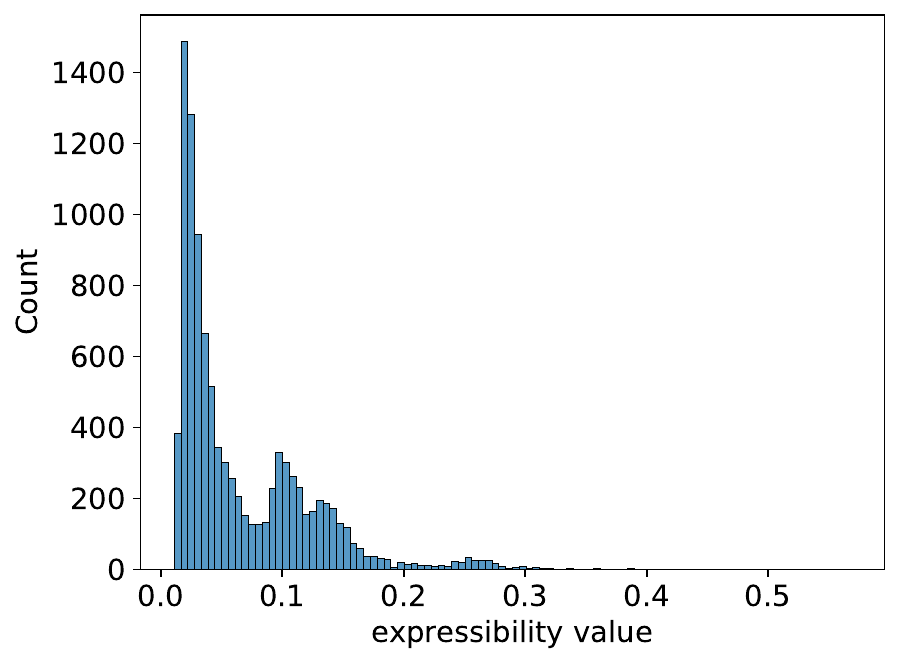}
        \caption{}
    \end{subfigure}
    \caption{Histograms depicting the number of paths and expressibility of generated circuits with 50 gates using the TeleGate method. (a) The number of paths for all $K_a$ circuits. (b) The expressibility of $K_p$ circuits filtered based on the number of paths.}
    \label{fig:path_express}
\end{figure}

Fig.\ \ref{fig:path_express} illustrates the distribution of the number of paths among the generated $K_a$ circuits, as well as the expressibility distribution of $K_p$ circuits featuring the highest number of paths. These circuits are generated using $N_g=50$ gates and the TeleGate method ($M$=TeleGate). It can be observed that the distribution of circuits is concentrated in the region with fewer paths, and as the number of paths increases, the distribution exhibits a long-tail phenomenon. Given the correlation between the number of paths and expressibility elucidated in Ref. \cite{he2024training-free}, we select the top $K_p$ circuits with the highest number of paths and calculate their expressibility. The distribution of expressibility among these selected circuits is concentrated in the region of low expressibility values, corresponding to better expressibility, thus ensuring promising performance in the VQE tasks. Similar distributions can be observed when varying $N_g$ and $M$.

\begin{figure}
	\centering
	\includegraphics[width = 18cm]{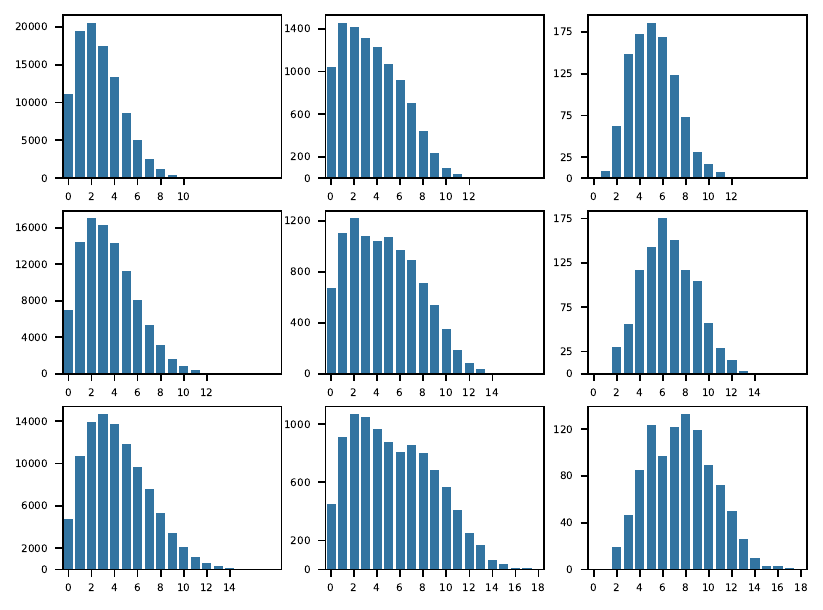}	
	\caption{Histograms depicting the number of ebits in generated circuits using the TeleGate method. The left, middle, and right column represent all $K_a$ circuits, $K_p$ circuits filtered by the number of paths, and $K_e$ circuits filtered by expressibility, respectively. The top, middle, and bottom row represent circuits with 40, 50, and 60 gates, respectively.}
	\label{fig:ebit}
\end{figure}

The distributions of the number of ebits for circuits generated with $N_g=40,50,60$ gates using the TeleGate method is depicted in Fig.\ \ref{fig:ebit}. The distributions approximate normal distributions. As the two-stage filtering process progresses, the mean of the normal distribution shifts towards higher values. A higher value of ebits corresponds to more nonlocal gates, thereby increasing the expressibility of the circuit. Similar distributions can be observed when employing the TeleData method or a combination of TeleGate and TeleData.

\begin{figure}[ht]
    \centering
    \begin{subfigure}[b]{0.32\textwidth}
        \centering
        \includegraphics[width=\textwidth]{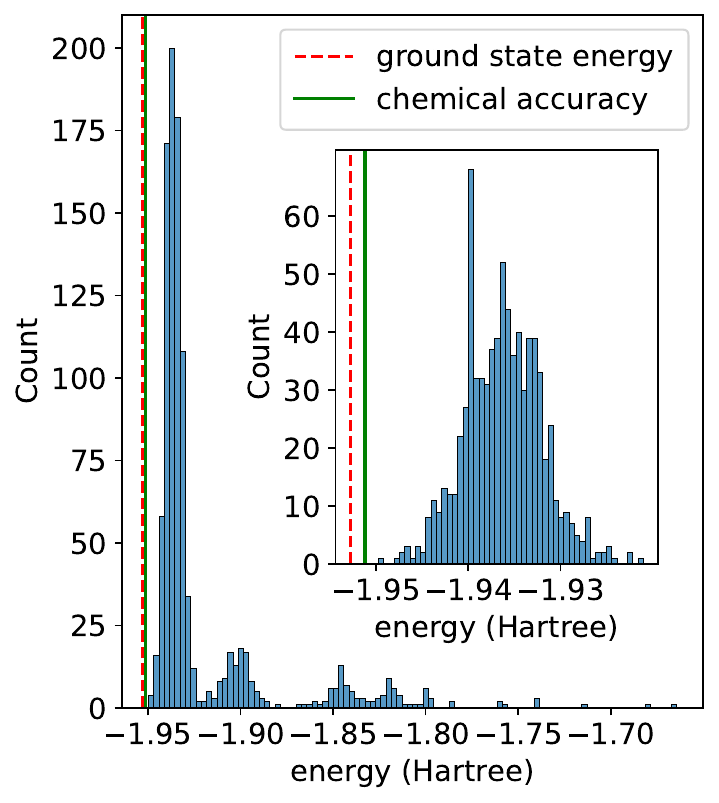}
        \caption{}
    \end{subfigure}
    \hfill
    \begin{subfigure}[b]{0.32\textwidth}
        \centering
        \includegraphics[width=\textwidth]{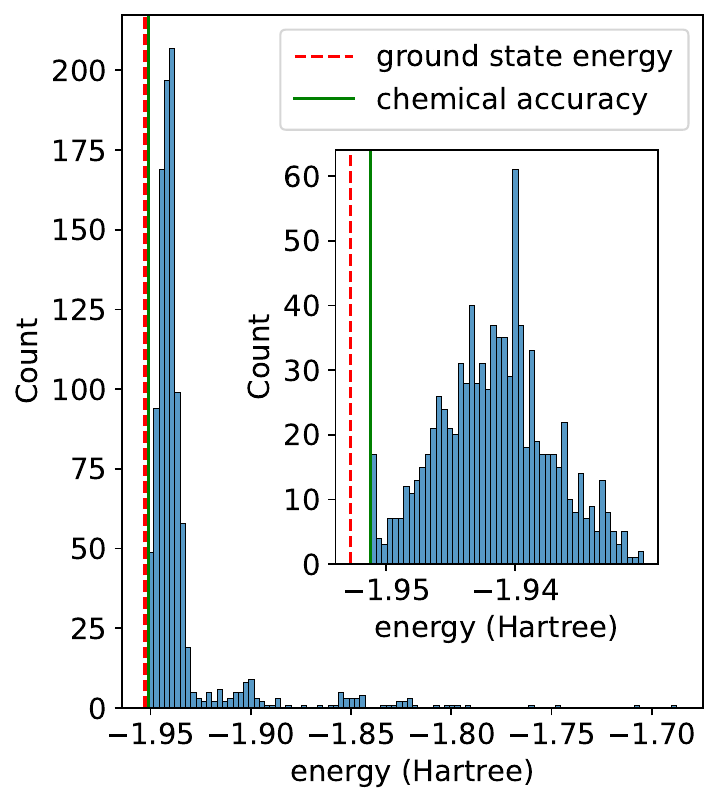}
        \caption{}
    \end{subfigure}
    \hfill
    \begin{subfigure}[b]{0.32\textwidth}
        \centering
        \includegraphics[width=\textwidth]{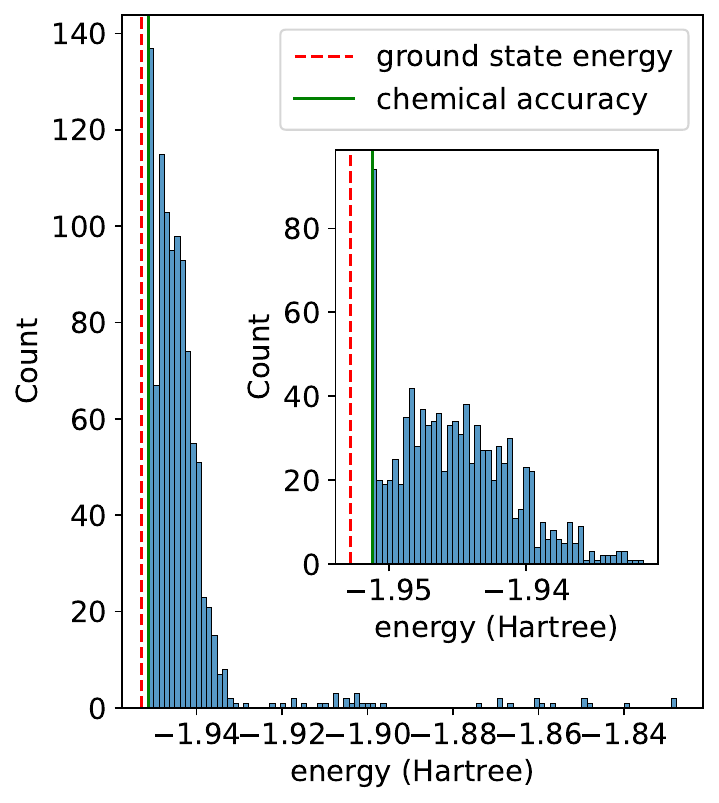}
        \caption{}
    \end{subfigure}
    \caption{Histograms depicting the lowest energy achieved for the BeH$_2$ problem by $K_e$ candidate circuits using the TeleGate method. Plots (a), (b), and (c) represent circuits with 40, 50, and 60 gates, respectively. The red dashed line denotes the ground state energy of BeH$_2$, while the green solid line indicates the energy within the chemical accuracy threshold of 0.0016.}
   	\label{fig:performance}
\end{figure}

After filtering the generated circuits by the number of paths and then the expressibility, the remaining circuits are queried for VQE tasks, following the order from better to worse expressibility. Fig.\ \ref{fig:performance} illustrates the achieved lowest energy for the BeH$_2$ problem among all $K_e$ candidate circuits when using the TeleGate method. As the number of gates increases, the distribution gradually shifts towards the optimal solution. With $N_g=40$, the gap between the found solution and the optimal one is 0.0030, slightly larger than the chemical accuracy threshold. As $N_g$ increases to 50, 13 optimal solutions are found, and this number rises to 79 when $N_g$ increases to 60.

\begin{figure}[ht]
    \centering
    \begin{subfigure}[b]{0.48\textwidth}
        \centering
        \includegraphics[width=\textwidth]{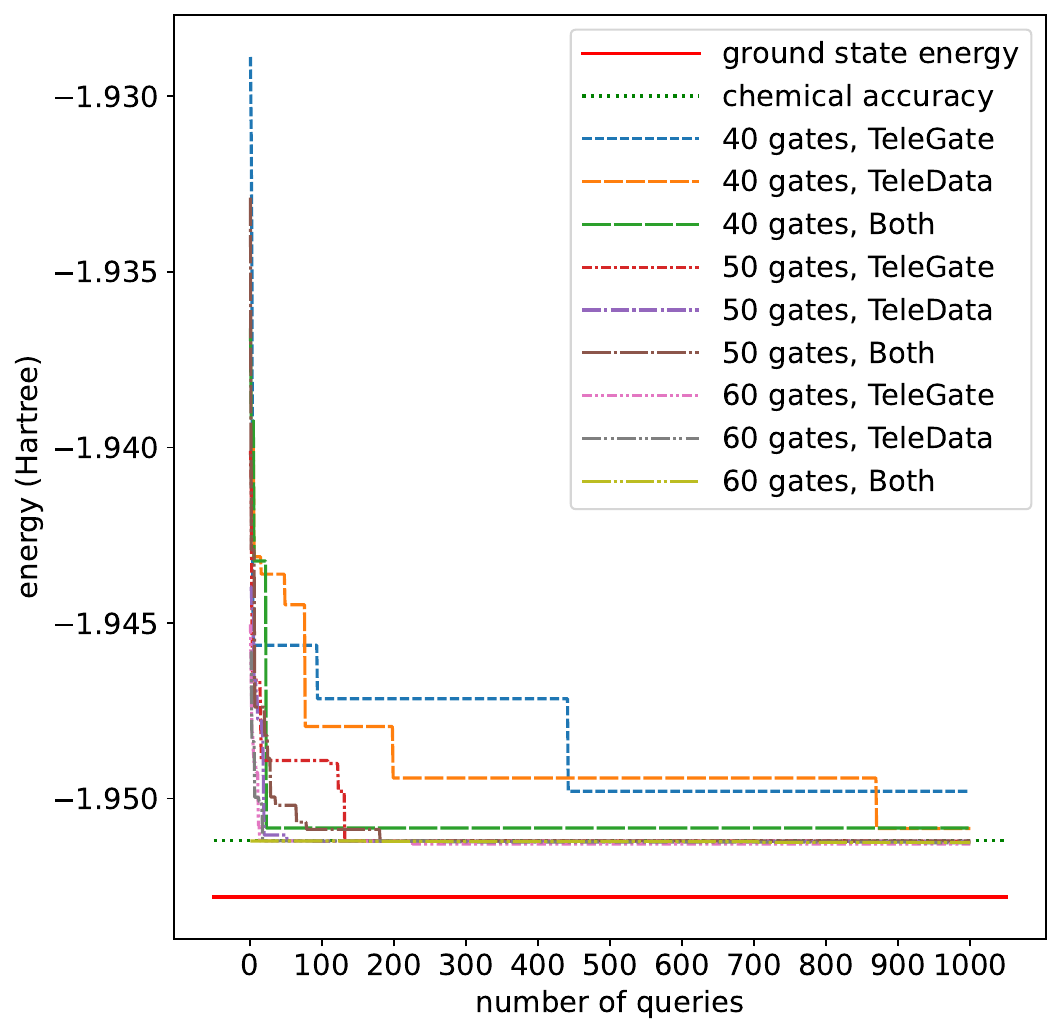}
        \caption{}
    \end{subfigure}
    \hfill
    \begin{subfigure}[b]{0.48\textwidth}
        \centering
        \includegraphics[width=\textwidth]{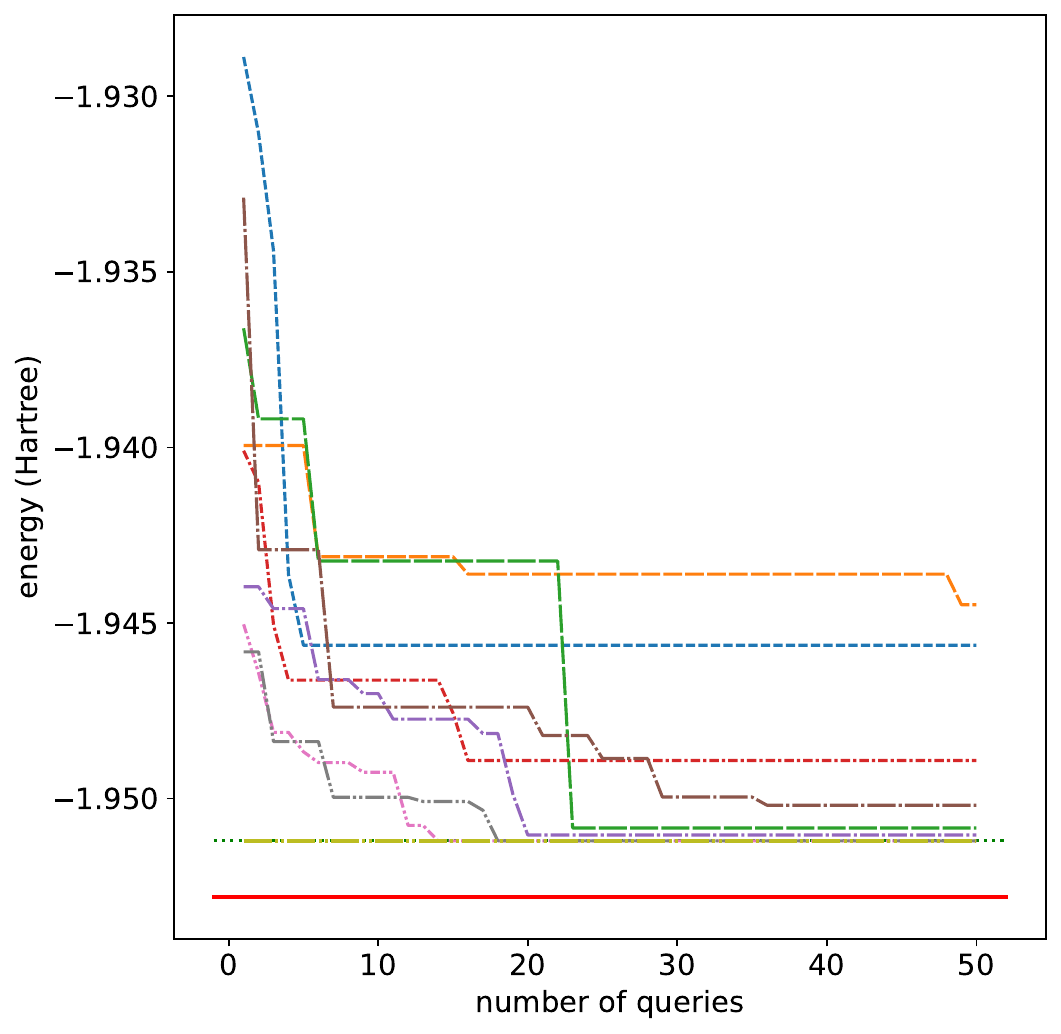}
        \caption{}
    \end{subfigure}
    \caption{Lowest energy achieved versus number of queries for BeH$_2$ problem. Figure (b) is a magnification of the first 50 queries in Figure (a).}
    \label{fig:query_energy}
\end{figure}

\begin{figure}
	\centering
	\includegraphics[width = 10cm]{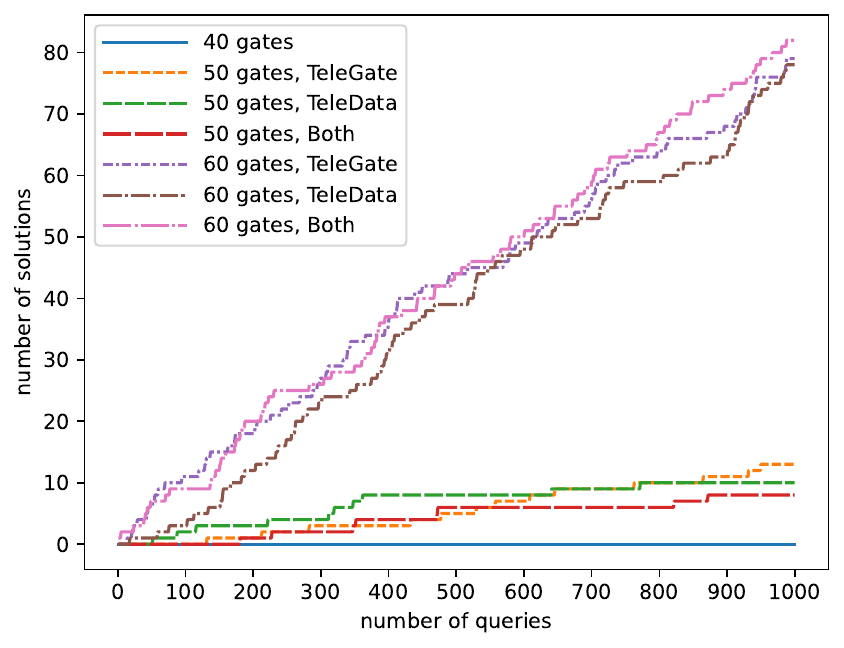}	
	\caption{Number of optimal solutions versus number of queries for BeH$_2$ problem.}
	\label{fig:query_sol}
\end{figure}

To further investigate the efficiency of finding optimal solutions, we depict the variation of the lowest energy achieved with the number of queries in Fig.\ \ref{fig:query_energy}. As more candidate circuits are queried, the lowest energy achieved decreases. Generally, circuits with more gates correspond to better solution and faster convergence to optimal solutions. For instance, with $N_g=60$ and $M$=Both, the first query successfully find the optimal solution. Similarly, the TeleGate and TeleData methods also discover the optimal solution within 20 queries. When $N_g$ decreases to 50, optimal solutions can be found within 200 queries. No optimal solution can be found when $N_g=40$. The overall query process is efficient as only a small number of candidate circuits need to be queried before finding the optimal solution. Fig.\ \ref{fig:query_sol} illustrates the relation between the number of optimal solutions and the number of queries. It can be seen that as the number of queries increases, there is a gradual increase in the number of optimal solutions.

Table \ref{tab:VQE_result} presents the simulation results for the 6-qubit BeH$_2$ molecule, Heisenberg model, and TFIM, respectively. It's evident that circuits comprising 40 gates fail to achieve the optimal solution. Notably, the gap between the minimal energy found and the ground state energy is already very close to the chemistry accuracy threshold of 0.0016 in the case of 6-qubit BeH$_2$ problem. Upon increasing the number of gates to 50, optimal solutions are achieved in all three problems with the combination of the TeleGate and TeleData method. Even when only TeleGate or TeleData is utilized and the optimal solution is not immediately found, the gap has significantly diminished. With the number of gates increasing to 60, using the TeleGate method and the TeleData method alone are capable of finding the optimal solutions, leading to an increase in the number of successful circuits. The minimal costs of ebits remains relatively consistent across different methods. For the BeH$_2$ problem using the TeleData method, and the Heisenberg problem using either the TeleGate or Both method, the consumed ebits decrease from 4 to 2 when the number of gates increases from 50 to 60. This points toward the possibility that increasing the number of gates may help discover circuits that consume fewer ebits.

\begin{table}
\centering
\begin{tabular}{ccccccccc}
\hline
& & \multicolumn{2}{c}{$N_g=40$}  & \multicolumn{3}{c}{$N_g=50$} & \multicolumn{2}{c}{$N_g=60$}\\
\hline
& & \#solution & gap & \#solution & gap & \#ebit& \#solution & \#ebit\\
\hline
& TeleGate & 0 & 0.0030 &  13 & - & 3 & 79 &  3\\
BeH$_2$& TeleData & 0 & 0.0019 &  10 & - & 4 & 78 &  2\\
& Both & 0 & 0.0020 &  8 & - & 3 & 82 &  2\\
\hline
& TeleGate & 0 & 0.2114 & 2 & - & 4 & 35 &  2\\
Heisenberg & TeleData & 0 & 0.3768 &  0 & 0.0095 & - & 29 &  3\\
& Both & 0 & 0.0720 &  2 & - & 4 & 35 &  2\\
\hline
& TeleGate & 0 & 0.0704 &  0 & 0.0128 & - & 5 &  3\\
TFIM & TeleData & 0 & 0.0647 &  0 & 0.0108 & - & 3 &  3\\
& Both     & 0 & 0.0464 &  1 & -      & 3 & 2 &  3\\
\hline
\end{tabular}
\caption{Simulation results for the 6-qubit BeH$_2$ molecule, Heisenberg model, and TFIM. ``\#solution'' indicates the number of candidate circuits that achieve the optimal solution. ``gap'' denotes the difference between the minimal energy found and the ground state energy. ``\#ebit'' indicates the minimum number of ebits used in the circuits that achieves the optimal solutions.}
\label{tab:VQE_result}
\end{table}

We also compare the results obtained from distributed QAS with those from the Hardware Effieicent Ansatz (HEA), commonly used in VQAs.
The HEA circuits are constructed by alternately placing odd and even layers. An odd layer is constructed as $V_{odd} = CNOT_{q_2}^{q_1}CNOT_{q_6}^{q_3}CNOT_{q_8}^{q_7}U_{q_1}U_{q_2}U_{q_3}U_{q_6}U_{q_7}U_{q_8}$, while an even layer is constructed as $V_{even} = CNOT_{q_3}^{q_2}CNOT_{q_7}^{q_6}U_{q_1}U_{q_2}U_{q_3}U_{q_6}U_{q_7}U_{q_8}$.
For instance, the circuit HEA-3 has 3 layers, constructed as $V_{odd}V_{even}V_{odd}$.
The nonlocal $CNOT_{q_6}^{q_3}$ can be implemented using the TeleGate method at the cost of 1 ebit. The comparison results in terms of the number of ebits, the number of parameters and the number of $CNOT$ gates are presented in Table \ref{tab:HEA_result}. For each problem, we first list the optimal circuits discovered by distributed QAS using the TeleGate, TeleData, and Both methods, followed by the optimal HEA circuit. The differences between the lowest energy achieved by these optimal circuits and the ground state energy are less than the chemical accuracy threshold of 0.0016. Additionally, the HEA circuit with the same number of ebits as the circuits found by distributed QAS and a comparable number of parameters is also listed, along with the gap between the lowest energy achieved and the ground state energy. From Table \ref{tab:HEA_result}, we can see that, compared to HEA circuits, the circuits found by distributed QAS achieve optimal solutions using only 40-60\% of the ebits, 44-57\% of the parameters, and a comparable number of $CNOT$ gates. This demonstrates that distributed QAS has a clear advantage in saving ebits, which are the most expensive resources in the context of distributed quantum computing.

\begin{table}
\centering
\begin{tabular}{cccccc}
\hline
& & \#ebit & \#parameter & \#$CNOT$ & gap\\
\hline
        & TeleGate & 3 &  72 & 19 & -\\
        & TeleData & 2 &  90 & 24 & -\\
BeH$_2$ & Both     & 2 &  90 & 25 & -\\
        & HEA-9    & 5 & 162 & 23 & -\\
        & HEA-6    & 3 & 108 & 15 & 0.0122\\
\hline
           & TeleGate & 2 &  84 & 26 & - \\
           & TeleData & 3 &  96 & 23 & -\\
Heisenberg & Both     & 2 &  90 & 24 & -\\
           & HEA-10   & 5 & 180 & 25 & -\\
           & HEA-6    & 3 & 108 & 15 & 0.3538\\
\hline
     & TeleGate & 3 &  93 & 24 & -\\
     & TeleData & 3 &  84 & 24 & -\\
TFIM & Both     & 3 &  78 & 20 & -\\
     & HEA-9    & 5 & 162 & 23 & -\\
     & HEA-6    & 3 & 108 & 15 & 0.0402\\
\hline
\end{tabular}
\caption{Comparison results between distributed QAS and HEA ansatz. HEA-$l$ indicates hardware efficient ansatz with $l$ layers. ``\#ebit'', ``\#parameter'', and ``\#CNOT''  indicate the number of ebits, the number of parameters, and the number of $CNOT$ gates used in the circuits, respectively. ``gap'' denotes the difference between the minimal energy found and the ground state energy. }
\label{tab:HEA_result}
\end{table}

\section{Discussion}\label{sec:discussion}

A major challenge in training PQCs is the barren plateau phenomenon \cite{larocca2024review}. When a PQC encounters a barren plateau, its optimization landscape becomes mostly flat as the size of the system increases. This causes the gradients to vanish exponentially, necessitating an exponentially large number of measurements to determine the parameter update direction. A fundamental relationship between expressibility and trainability has been derived \cite{holmes2022connecting}, indicating that highly expressive ansatz exhibit flatter cost landscapes and therefore will be harder to train. In the proposed distributed QAS, the two-stage filtering process chooses circuits with best expressibility. These circuits are then trained to optimize the loss function of the specific problem. However, due to the impact of barren plateaus, circuits with poor trainability may converge to non-optimal solutions or have their training suspended once the number of training iterations reaches a predefined threshold. To mitigate this, we use multiple initializations of parameters when training a PQC, similar to the approach taken in \cite{zhang2021neural}, thus increasing the likelihood that at least one initialization will avoid barren plateaus. A more effective solution is to consider trainability during the filtering process. Therefore, in future work, we plan to design a training-free proxy capable of reflecting the trainability of a PQC. This will help filter out circuits with poor trainability, thereby avoiding the waste of time on training such circuits.

Our distributed QAS framework adopts a training-free approach, utilizing low-computation-cost metrics to evaluate quantum circuits. This approach enables us to explore a greater number of quantum circuits compared to training-based methods, thereby increasing the likelihood of finding optimal circuits.
To better address larger-scale quantum circuit search problems, we can incorporate elements of the VAns method \cite{bilkis2023semi} into our framework. Specifically, by following the gate insertion rule of VAns, we can add blocks of gates to expand the circuit during the quantum circuit generation process, rather than inserting one quantum gate at a time. This enables the generation of larger-scale quantum circuits without increasing computational costs. Additionally, placing blocks of quantum gates reduces the need to select qubit positions for each individual gate, thereby controlling the size of the search space and preventing exponential growth in the required number of circuit samples as the number of qubits increases. Meanwhile, the gate simplification rule of VAns can be employed to remove unnecessary or redundant gates, or gates with minimal impact on cost, thereby managing the growth in the number of gates.

Two commonly used nonlocal gate implementation methods are incorporated into our distributed QAS framework. In general, the TeleGate method may benefit the effective utilization of ebits. When a qubit enters control mode, multiple nonlocal or local CNOT gates using it as the control can be implemented, until the qubit becomes the target of certain gate and exits control mode. On the other hand, the TeleData method may reduce the number of nonlocal gates. After a qubit is teleported to another QPU, all gates acting on it and the native qubits of that QPU become local gates. Allowing both methods increases the flexibility of distributed quantum circuits. As shown in Table \ref{tab:HEA_result}, for each problem, the ebit consumption using both methods is no more than the consumption when using only the TeleGate method or only the TeleData method.

\section{Conclusion}\label{sec:conclusion}
In distributed quantum computing, the entire distributed system is composed of all qubits of multiple interconnected QPUs, and nonlocal gates across different QPUs can be implemented by methods like TeleGate or TeleData. Although various aspects of QAS have been investigated in recent years, how to design distributed quantum circuit structures automatically remained unexplored. This problem poses a more complex challenge because it requires optimization not only of gate types and positions, but also of the implementation method for nonlocal gates.

In this work, we have proposed a distributed QAS framework for multiple interconnected QPUs with specific qubit connectivity. The integration of TeleGate and TeleData, as well as the qubit assignment from logical to physical qubits, makes our QAS framework very flexible in exploring diverse circuit structures. The training-free evaluation methodology also benefits the exploration of huge amount of quantum structures, enhancing the likelihood of discovering resource-efficient circuits. Considering the qubit connectivity of two IBM quantum processors, we use the proposed framework to find distributed quantum circuits for computing the ground state energy of the BeH$_2$ molecule, the Heisenberg model and the transverse-field Ising model. The optimal solutions have been achieved for these problems, although only a small set of circuits are trained during the entire QAS workflow.


\section*{Acknowledgements}
This work is supported by Guangdong Basic and Applied Basic Research Foundation (Nos. 2022A1515140116, 2021A1515011985), Jihua Laboratory Scientific Project (No. X210101UZ210), Innovation Program for Quantum Science and Technology (No. 2021ZD0302901), Guangdong Provincial Quantum Science Strategic Initiative (No. GDZX2303007), and the National Natural Science Foundation of China (Nos. 62272492).

\section*{Appendix A}
In this Appendix, we provide a detailed description of the cat-entangler and cat-disentangler primitive operations \cite{yimsiriwattana2004generalized}, which support the TeleGate method described in Section \ref{sec:TeleGate}.

The quantum state before the cat-entangler operation is
\begin{align}
|\psi\rangle(|00\rangle+|11\rangle)_{ab} = (|0\rangle_c|\psi_0\rangle + |1\rangle_c|\psi_1\rangle)(|00\rangle+|11\rangle)_{ab}.
\end{align}
The cat-entangler comprises  (1) a $CNOT^c_a$, (2) a measurement on qubit $a$, and (3) a conditional $X$ gate on qubit $b$ based on the measurement outcome. Following the first step, the quantum state transforms to
\begin{align}
CNOT^c_a|\psi\rangle (|00\rangle+|11\rangle)_{ab}
& = |0\rangle_c|\psi_0\rangle(|00\rangle+|11\rangle)_{ab} + |1\rangle_c|\psi_1\rangle(|10\rangle+|01\rangle)_{ab}.
\end{align}
Afterwards, qubit $a$ is measured in the computational basis. If the measurement outcome is 0, the state collapses to
\begin{align}
|00\rangle_{cb}|\psi_0\rangle + |11\rangle_{cb}|\psi_1\rangle,
\end{align}
otherwise it collapses to
\begin{align}
|01\rangle_{cb}|\psi_0\rangle + |10\rangle_{cb}|\psi_1\rangle.
\end{align}
In the third step, a $X$ gate is applied to qubit $b$ only if the measurement outcome is 1. Therefore, regardless of the measurement outcome, the state after the cat-entangler operation is always
\begin{align}
|00\rangle_{cb}|\psi_0\rangle + |11\rangle_{cb}|\psi_1\rangle.
\end{align}

The cat-disentangler comprises (1) a Hadamard gate $H$ on qubit $b$, (2) a measurement on qubit $b$, and (3) a conditional $Z$ gate on qubit $c$ based on the measurement outcome. After the first step, the quantum state becomes
\begin{align}
|0\rangle_{c}(|0\rangle+|1\rangle)_b|\psi_0\rangle + |1\rangle_{c}(|0\rangle-|1\rangle)_b|\psi_1\rangle.
\end{align}
Then qubit $b$ is measured in the computational basis. If the measurement outcome is 0, the state collapses to
\begin{align}
|0\rangle_{c}|\psi_0\rangle + |1\rangle_{c}|\psi_1\rangle,
\end{align}
otherwise it collapses to
\begin{align}
|0\rangle_{c}|\psi_0\rangle - |1\rangle_{c}|\psi_1\rangle.
\end{align}
In the third step, a $Z$ gate is performed on qubit $c$ only if the measurement outcome is 1. Thus, regardless of the measurement outcome, the state after the cat-disentangler is always
\begin{align}
|0\rangle_{c}|\psi_0\rangle + |1\rangle_{c}|\psi_1\rangle.
\end{align}

\bibliographystyle{unsrt}
\bibliography{Reference}

\begin{thebibliography}{10}

\bibitem{cerezo2021variational}
Marco Cerezo, Andrew Arrasmith, Ryan Babbush, Simon~C Benjamin, Suguru Endo,
  Keisuke Fujii, Jarrod~R McClean, Kosuke Mitarai, Xiao Yuan, Lukasz Cincio,
  and Patrick~J Coles.
\newblock Variational quantum algorithms.
\newblock {\em Nature Reviews Physics}, 3(9):625--644, 2021.

\bibitem{liang2021hybrid}
Yanying Liang, Wei Peng, Zhu-Jun Zheng, Olli Silv{\'e}n, and Guoying Zhao.
\newblock A hybrid quantum--classical neural network with deep residual
  learning.
\newblock {\em Neural Networks}, 143:133--147, 2021.

\bibitem{situ2020quantum}
Haozhen Situ, Zhimin He, Yuyi Wang, Lvzhou Li, and Shenggen Zheng.
\newblock Quantum generative adversarial network for generating discrete
  distribution.
\newblock {\em Information Sciences}, 538:193--208, 2020.

\bibitem{zhou2023hybrid}
Nan-Run Zhou, Tian-Feng Zhang, Xin-Wen Xie, and Jun-Yun Wu.
\newblock Hybrid quantum--classical generative adversarial networks for image
  generation via learning discrete distribution.
\newblock {\em Signal Processing: Image Communication}, 110:116891, 2023.

\bibitem{wang2023quantum}
Yizhi Wang, Shichuan Xue, Yaxuan Wang, Yong Liu, Jiangfang Ding, Weixu Shi,
  Dongyang Wang, Yingwen Liu, Xiang Fu, Guangyao Huang, Anqi Huang, Mingtang
  Deng, and Junjie Wu.
\newblock Quantum generative adversarial learning in photonics.
\newblock {\em Optics Letters}, 48(20):5197--5200, 2023.

\bibitem{zhao2022qsan}
Jinjing Shi, Ren-Xin Zhao, Wenxuan Wang, Shichao Zhang, and Xuelong Li.
\newblock {QSAN}: A near-term achievable quantum self-attention network.
\newblock {\em arXiv preprint arXiv:2207.07563}, 2022.

\bibitem{zhao2023qksan}
Ren-Xin Zhao, Jinjing Shi, and Xuelong Li.
\newblock {QKSAN}: A quantum kernel self-attention network.
\newblock {\em arXiv preprint arXiv:2308.13422}, 2023.

\bibitem{zhao2024gqhan}
Ren-Xin Zhao, Jinjing Shi, and Xuelong Li.
\newblock {GQHAN}: A {G}rover-inspired quantum hard attention network.
\newblock {\em arXiv preprint arXiv:2401.14089}, 2024.

\bibitem{parigi2023quantum}
Marco Parigi, Stefano Martina, and Filippo Caruso.
\newblock Quantum-noise-driven generative diffusion models.
\newblock {\em arXiv preprint arXiv:2308.12013}, 2023.

\bibitem{zhang2024generative}
Bingzhi Zhang, Peng Xu, Xiaohui Chen, and Quntao Zhuang.
\newblock Generative quantum machine learning via denoising diffusion
  probabilistic models.
\newblock {\em Physical Review Letters}, 132(10):100602, 2024.

\bibitem{chen2024quantum}
Chuangtao Chen, Qinglin Zhao, Mengchu Zhou, Zhimin He, Zhili Sun, and Haozhen
  Situ.
\newblock Quantum generative diffusion model: A fully quantum-mechanical model
  for generating quantum state ensemble.
\newblock {\em arXiv preprint arXiv:2401.07039}, 2024.

\bibitem{liu2023quantum}
Zidu Liu, Pei-Xin Shen, Weikang Li, Lu-Ming Duan, and Dong-Ling Deng.
\newblock Quantum capsule networks.
\newblock {\em Quantum Science and Technology}, 8(1):015016, 2023.

\bibitem{pan2023deep}
Xiaoxuan Pan, Zhide Lu, Weiting Wang, Ziyue Hua, Yifang Xu, Weikang Li, Weizhou
  Cai, Xuegang Li, Haiyan Wang, Yi-Pu Song, Chang-Ling Zou, Dong-Ling Deng, and
  Luyan Sun.
\newblock Deep quantum neural networks on a superconducting processor.
\newblock {\em Nature Communications}, 14:4006, 2023.

\bibitem{herrmann2022realizing}
Johannes Herrmann, Sergi~Masot Llima, Ants Remm, Petr Zapletal, Nathan~A
  McMahon, Colin Scarato, Fran{\c{c}}ois Swiadek, Christian~Kraglund Andersen,
  Christoph Hellings, Sebastian Krinner, Nathan Lacroix, Stefania Lazar,
  Michael Kerschbaum, Dante~Colao Zanuz, Graham~J Norris, Michael~J Hartmann,
  Andreas Wallraff, and Christopher Eichler.
\newblock Realizing quantum convolutional neural networks on a superconducting
  quantum processor to recognize quantum phases.
\newblock {\em Nature Communications}, 13:4144, 2022.

\bibitem{ren2021comprehensive}
Pengzhen Ren, Yun Xiao, Xiaojun Chang, Po-Yao Huang, Zhihui Li, Xiaojiang Chen,
  and Xin Wang.
\newblock A comprehensive survey of neural architecture search: Challenges and
  solutions.
\newblock {\em ACM Computing Surveys (CSUR)}, 54(4):1--34, 2021.

\bibitem{cincio2018learning}
Lukasz Cincio, Yi{\u{g}}it Suba{\c{s}}{\i}, Andrew~T Sornborger, and Patrick~J
  Coles.
\newblock Learning the quantum algorithm for state overlap.
\newblock {\em New Journal of Physics}, 20(11):113022, 2018.

\bibitem{lu2021markovian}
Zhide Lu, Pei-Xin Shen, and Dong-Ling Deng.
\newblock Markovian quantum neuroevolution for machine learning.
\newblock {\em Physical Review Applied}, 16(4):044039, 2021.

\bibitem{meng2021quantum}
Fan-Xu Meng, Ze-Tong Li, Xu-Tao Yu, and Zai-Chen Zhang.
\newblock Quantum circuit architecture optimization for variational quantum
  eigensolver via {M}onto {C}arlo tree search.
\newblock {\em IEEE Transactions on Quantum Engineering}, 2:1--10, 2021.

\bibitem{altares2021automatic}
Sergio Altares-L{\'o}pez, Angela Ribeiro, and Juan~Jos{\'e} Garc{\'\i}a-Ripoll.
\newblock Automatic design of quantum feature maps.
\newblock {\em Quantum Science and Technology}, 6(4):045015, 2021.

\bibitem{huang2022robust}
Yuhan Huang, Qingyu Li, Xiaokai Hou, Rebing Wu, Man-Hong Yung, Abolfazl Bayat,
  and Xiaoting Wang.
\newblock Robust resource-efficient quantum variational ansatz through an
  evolutionary algorithm.
\newblock {\em Physical Review A}, 105(5):052414, 2022.

\bibitem{zhang2021neural}
Shi-Xin Zhang, Chang-Yu Hsieh, Shengyu Zhang, and Hong Yao.
\newblock Neural predictor based quantum architecture search.
\newblock {\em Machine Learning: Science and Technology}, 2(4):045027, 2021.

\bibitem{du2022quantum}
Yuxuan Du, Tao Huang, Shan You, Min-Hsiu Hsieh, and Dacheng Tao.
\newblock Quantum circuit architecture search for variational quantum
  algorithms.
\newblock {\em npj Quantum Information}, 8(1):62, 2022.

\bibitem{caleffi2022distributed}
Marcello Caleffi, Michele Amoretti, Davide Ferrari, Daniele Cuomo, Jessica
  Illiano, Antonio Manzalini, and Angela~Sara Cacciapuoti.
\newblock Distributed quantum computing: a survey.
\newblock {\em arXiv preprint arXiv:2212.10609}, 2022.

\bibitem{he2024training-free}
Zhimin He, Maijie Deng, Shenggen Zheng, Lvzhou Li, and Haozhen Situ.
\newblock Training-free quantum architecture search.
\newblock In {\em Proceedings of the AAAI Conference on Artificial
  Intelligence}, volume~38, pages 12430--12438, 2024.

\bibitem{bilkis2023semi}
M~Bilkis, M~Cerezo, Guillaume Verdon, Patrick~J Coles, and Lukasz Cincio.
\newblock A semi-agnostic ansatz with variable structure for variational
  quantum algorithms.
\newblock {\em Quantum Machine Intelligence}, 5(2):43, 2023.

\bibitem{grimsley2019adaptive}
Harper~R Grimsley, Sophia~E Economou, Edwin Barnes, and Nicholas~J Mayhall.
\newblock An adaptive variational algorithm for exact molecular simulations on
  a quantum computer.
\newblock {\em Nature Communications}, 10:3007, 2019.

\bibitem{he2023gsqas}
Zhimin He, Maijie Deng, Shenggen Zheng, Lvzhou Li, and Haozhen Situ.
\newblock {GSQAS}: Graph self-supervised quantum architecture search.
\newblock {\em Physica A: Statistical Mechanics and its Applications},
  630:129286, 2023.

\bibitem{li2023eqnas}
Yangyang Li, Ruijiao Liu, Xiaobin Hao, Ronghua Shang, Peixiang Zhao, and
  Licheng Jiao.
\newblock {EQNAS}: Evolutionary quantum neural architecture search for image
  classification.
\newblock {\em Neural Networks}, 168:471--483, 2023.

\bibitem{wang2023automated}
Peiyong Wang, Muhammad Usman, Udaya Parampalli, Lloyd~CL Hollenberg, and
  Casey~R Myers.
\newblock Automated quantum circuit design with nested {M}onte {C}arlo tree
  search.
\newblock {\em IEEE Transactions on Quantum Engineering}, 2023.

\bibitem{wang2022quantumnas}
Hanrui Wang, Yongshan Ding, Jiaqi Gu, Yujun Lin, David~Z Pan, Frederic~T Chong,
  and Song Han.
\newblock Quantum{NAS}: Noise-adaptive search for robust quantum circuits.
\newblock In {\em 2022 IEEE International Symposium on High-Performance
  Computer Architecture (HPCA)}, pages 692--708. IEEE, 2022.

\bibitem{zhang2022differentiable}
Shi-Xin Zhang, Chang-Yu Hsieh, Shengyu Zhang, and Hong Yao.
\newblock Differentiable quantum architecture search.
\newblock {\em Quantum Science and Technology}, 7(4):045023, 2022.

\bibitem{wu2023quantumdarts}
Wenjie Wu, Ge~Yan, Xudong Lu, Kaisen Pan, and Junchi Yan.
\newblock Quantum{DARTS}: Differentiable quantum architecture search for
  variational quantum algorithms.
\newblock In {\em International Conference on Machine Learning}, pages
  37745--37764. PMLR, 2023.

\bibitem{he2022quantum}
Zhimin He, Chuangtao Chen, Lvzhou Li, Shenggen Zheng, and Haozhen Situ.
\newblock Quantum architecture search with meta-learning.
\newblock {\em Advanced Quantum Technologies}, 5(8):2100134, 2022.

\bibitem{yimsiriwattana2004generalized}
Anocha Yimsiriwattana and Samuel~J Lomonaco~Jr.
\newblock Generalized {GHZ} states and distributed quantum computing.
\newblock {\em arXiv preprint quant-ph/0402148}, 2004.

\bibitem{bennett1993teleporting}
Charles~H Bennett, Gilles Brassard, Claude Cr{\'e}peau, Richard Jozsa, Asher
  Peres, and William~K Wootters.
\newblock Teleporting an unknown quantum state via dual classical and
  {E}instein-{P}odolsky-{R}osen channels.
\newblock {\em Physical Review Letters}, 70(13):1895, 1993.

\bibitem{sim2019expressibility}
Sukin Sim, Peter~D Johnson, and Al{\'a}n Aspuru-Guzik.
\newblock Expressibility and entangling capability of parameterized quantum
  circuits for hybrid quantum-classical algorithms.
\newblock {\em Advanced Quantum Technologies}, 2(12):1900070, 2019.

\bibitem{kandala2017hardware}
Abhinav Kandala, Antonio Mezzacapo, Kristan Temme, Maika Takita, Markus Brink,
  Jerry~M Chow, and Jay~M Gambetta.
\newblock Hardware-efficient variational quantum eigensolver for small
  molecules and quantum magnets.
\newblock {\em Nature}, 549(7671):242--246, 2017.

\bibitem{zhang2023tensorcircuit}
Shi-Xin Zhang, Jonathan Allcock, Zhou-Quan Wan, Shuo Liu, Jiace Sun, Hao Yu,
  Xing-Han Yang, Jiezhong Qiu, Zhaofeng Ye, Yu-Qin Chen, Chee-Kong Lee, Yi-Cong
  Zheng, Shao-Kai Jian, Hong Yao, Chang-Yu Hsieh, and Shengyu Zhang.
\newblock Tensor{C}ircuit: a quantum software framework for the {NISQ} era.
\newblock {\em Quantum}, 7:912, 2023.

\bibitem{larocca2024review}
Mart\'{\i}n Larocca, Supanut Thanasilp, Samson Wang, Kunal Sharma, Jacob
  Biamonte, Patrick~J Coles, Lukasz Cincio, Jarrod~R McClean, Zo\"{e} Holmes,
  and M~Cerezo.
\newblock A review of barren plateaus in variational quantum computing.
\newblock {\em arXiv preprint arXiv:2405.00781}, 2024.

\bibitem{holmes2022connecting}
Zo{\"e} Holmes, Kunal Sharma, M~Cerezo, and Patrick~J Coles.
\newblock Connecting ansatz expressibility to gradient magnitudes and barren
  plateaus.
\newblock {\em PRX Quantum}, 3(1):010313, 2022.

\end{thebibliography}
\end{document}